\documentclass[pra,twocolumn,showpacs,superscriptaddress]{revtex4}

\usepackage{amsmath}    
\usepackage{graphicx}   
\usepackage{verbatim}   
\usepackage{color}      
\usepackage{subfigure}  
\usepackage{hyperref}   

\begin{document}

\title{Laser-assisted-autoionization dynamics of helium resonances \\with single attosecond pulses}
\date{\today}
\author{Wei-Chun Chu}
\affiliation{J. R. Macdonald Laboratory, Department of Physics, Kansas State University, Manhattan, Kansas 66506, USA}
\author{Song-Feng Zhao}
\affiliation{Key Laboratory of Atomic and Molecular Physics and Functional Materials of Gansu Province,
College of Physics and Electronic Engineering, Northwest Normal University, Lanzhou 730070, China}
\author{C. D. Lin}
\affiliation{J. R. Macdonald Laboratory, Department of Physics, Kansas State University, Manhattan, Kansas 66506, USA}

\begin{abstract}
The strong coupling between two autoionizing states in helium is studied theoretically with the pump-probe scheme. An isolated
100-attosecond XUV pulse is used to excite helium near the $2s2p(^1P)$ resonance state in the presence of an intense infrared (IR)
laser. The laser field introduces strong coupling between $2s2p(^1P)$ and $2p^2(^1S)$ states. The IR also can ionize helium from both
autoionizing states. By changing the time delay between the XUV and the IR pulses, we investigated the photoelectron spectra near the
two resonances. The results are used to explain the recent experiment by Gilbertson \textit{et al}
[Phys. Rev. Lett. 105, 263003 (2010)]. Using the
same isolated attosecond pulse and a 540 nm laser, we also investigate the strong coupling between $2s2p(^1P)$ and $2s^2(^1S)$
by examining how the photoelectron spectra are modified vs the time delay and the possibility of observing Autler-Townes doublet in
such experiments.
\end{abstract}

\maketitle

\section{Introduction}\label{intro}

Quantum coherence is essential to the understanding and control of the dynamics in a quantum system. With the advances of
experimental techniques and technologies, extreme ultraviolet (XUV) attosecond pulse trains (APT) and single attosecond pulses (SAP)
have been produced through the high-order harmonic generation (HHG) process, by exposing atoms to intense infrared lasers. Since the
natural time scale of the electronic motion of valence electrons in atoms and molecules is in the order of hundreds of attoseconds,
these pulses can be used to manipulate or control electron dynamics at the attosecond level~\cite{krausz}. For this purpose,
pump-probe scheme is used experimentally where the dynamics of the system is investigated by varying the time delay between the
pulses. However, existing XUV SAP and APT are too feeble to be used for XUV-pump-XUV-probe measurements. Instead, most of the
experiments have been carried out using the weak XUV-pump with an intense IR-probe pulse. Rigorously speaking, the IR can be
considered as a probe only if it is applied after the XUV pump ends. In such case, the electron wave packet generated by the APT or
the SAP is probed. In most experiments, however, measurements are also made where the XUV and IR
overlap in time. Since the IR is much stronger than the XUV, such measurements are better understood as laser-assisted
photoionization.

In XUV+IR experiments, the interaction of the XUV+IR field with the target atom is a nonlinear process. To treat such problems
theoretically, the brute-force numerical solution of the time-dependent Schr\"{o}dinger equation (TDSE) has been carried out by many
groups where the target atom is treated within the single electron approximation. For APT+IR, Floquet approach~\cite{tong10}
has also been employed as well. With much efforts, TDSE calculations have been performed for two-electron helium atom and
two-electron hydrogen molecules and compared with measurements~\cite{kelkensberg-sap, kelkensberg-apt, sansone}.
Still, simplified theoretical models are highly desirable in order to acquire a better understanding of the dynamics.
These models, clearly, have to depend on the physical systems on hand. To be specific, we will focus on helium atom where many
experiments have been carried out.

The first type of XUV+IR experiments used APT where the photon energy runs from below to above the single ionization threshold of
helium~\cite{johnsson, ranitovic10, ranitovic11, mauritsson, shivaram}. Spectra of electrons or ions have been measured and
calculations based on TDSE and Floquet theory reported~\cite{johnsson, ranitovic11}. In the limit of non-overlapping APT
and IR, a simplified theory has been reported where the aim was to extract the electron wave packet generated by
the APT~\cite{choi}. Alternatively, photon spectra can also be measured~\cite{holler}. In this case, effect due to propagation in the
medium has to be considered~\cite{gaarde}. As the energy of the XUV photon increases beyond the single-ionization threshold, it
enters a structureless spectral region (about 25 to 55 eV). In this case, the addition of an intense IR to the XUV is to shift the
momentum of the continuum electron along the polarization direction of the IR laser, resulting in the "streaking" of photoelectrons
where the energies depend on the vector potential of the IR laser at the time of the electron emission. A "streaking" theory based
on the strong field approximation (SFA)~\cite{kitzler} has been widely used. This theory also forms the basis of the
frequency-resolved optical gating for complete reconstruction of attosecond bursts method~\cite{mairesse} for extracting the pulse
duration of an SAP. At a still higher photon energy of 60 eV, the XUV alone will
reach the spectral region where doubly-excited states of helium are located. In this paper, we will focus on this energy region and
consider the dynamics of the doubly excited state generated by the XUV pulse in the presence of an IR pulse at different delay
times.

Doubly excited states have been widely studied using synchrotron radiations since the 1960's~\cite{madden, domke91}. These are
autoionizing states where the lifetime of a state is typically of a few to tens of femtoseconds. These lifetimes are deduced from the
measured spectral width; thus high-precision spectroscopic measurement is needed. With the emergence of SAP, the lifetime of an Auger
resonance was first measured in the time domain and analyzed by the XUV+IR streaking theory in 2002~\cite{drescher}. However, the
evolution of the spectral shape of a resonance state has not been determined so far in the time-domain measurements, which in
principle can be investigated with attosecond pulses. In fact, the autoionization theory was formulated in the energy domain by
Fano~\cite{fano}. This issue was addressed recently by us~\cite{chu} and an experimental scheme was proposed. The scheme, however,
requires an attosecond XUV-pump to create the resonance and an attosecond XUV-probe to project out the time-evolution of the
resonance. Such measurement is not possible yet. Thus for the time domain measurements so far, doubly-excited states still have to be
"probed" using an IR pulse. For an isolated Fano resonance generated in a combined XUV+IR pulse, the problem has been examined by
Wickenhauser \textit{et al}~\cite{wickenhauser}. A simpler model was proposed by Zhao \textit{et al}~\cite{zhao} based on SFA.

An XUV+IR experiment on helium doubly-excited states has been reported by Loh \textit{et al}~\cite{loh}, where a 30 fs XUV pulse was
used to excite the $2s2p(^1P)$ doubly excited state in the presence of a 42 fs IR laser. The absorption spectra near the $2s2p$
resonance were measured vs the time delay between the pulses. The result was then interpreted by a theory that includes the strong
coupling between the $2s2p$ and $2p^2(^1S)$ states by the IR. In this experiment, the pulse duration of the XUV was longer than the
lifetime of the $2s2p$ autoionizing state, i.e., the bandwidth of the light was narrower than the resonance width. Thus, the 
continuum electrons in a measurement did not have a meaningful ``distribution'' for such a narrow bandwidth. A single pump could not
reveal any significant spectroscopic features; instead, sequential measurements were made where the detuning of the XUV was
controlled and scanned through an energy range. The scanned spectra were able to exhibit the autoionization modified by the
time-delayed IR, thus providing the temporal information of the system  partially. More recently, an 100 as SAP was used by
Gilbertson \textit{et al}~\cite{gilbertson} to excite helium in the presence of a 9 fs IR pulse. The central energy of the excitation 
is at the $2s2p$ resonance, and the photoelectron spectra near this resonance vs time delay were reported. The result was
interpreted based on the theory of Zhao \textit{et al}~\cite{zhao} from which the decay lifetime of $2s2p$ was retrieved. However,
the spectral shape of the resonance was not analyzed due to the limited resolution of 0.7 eV of the electron
spectrometer. In contrast to Ref.~\cite{loh}, the interpretation was made by the streaking model of Zhao \textit{et al}~\cite{zhao} 
only. Since the IR wavelength used in the two experiments are about the same, the role of strong coupling between the $2s2p$ and
$2p^2$ states in this experiment should be addressed. This is the goal of the present paper.

The present analysis limits the Hilbert space to including only the two autoionizing states, $2s2p$ and $2p^2$, and the ground state. 
While the XUV can populate other higher doubly excited states, they are weaker in the spectrum and are easily ionized by the IR.
Governed by the detuning of the 9 fs IR, only the $2s2p$ and $2p^2$ states are strongly coupled. The dynamics of such a system is
then studied with a theory generalized from the standard three-level system formulated earlier~\cite{lambropoulos, bachau, madsen,
themelis}. To be specific, we develop a model for the IR-dressed autoionization of the $2s2p$ resonance in helium excited by an SAP,
where the IR strongly couples the $2s2p$ and $2p^2$ resonances and also ionizes both doubly-excited states. Both effects by the IR 
have to be included in order to explain the observed electron spectra reported in Ref.~\cite{gilbertson}. After achieving good
agreement with the experiment, we further investigate the case where $2s2p$ and $2s^2(^1S)$ are coupled by changing the laser
wavelength to 540~nm. The latter state has a higher binding energy such that it is not ionized by the laser. Such measurements would
be closer to the standard electromagnetically induced transparency (EIT) experiment~\cite{harris, fleischhauer}. We will look for the
presence of the Autler-Townes doublet~\cite{autler} which are routinely generated by three-level systems with long pulses.

In Sec.~\ref{theory}, the model system is defined and the method is introduced. In Sec.~\ref{results}, we show and analyze the 
results of the two cases, with laser wavelengths $\lambda_L=780$ and 540~nm, respectively. For $\lambda_L=780$~nm, our calculation is
compared with the available experiment~\cite{gilbertson} and with the streaking model based on the SFA model of Zhao \textit{et
al}~\cite{zhao}. For $\lambda_L=540$~nm, the results are analyzed, and the possible experimental realization are discussed. In
Sec.~\ref{conclusions}, we give the conclusion. Atomic units (a.u.) are used in Sec.~\ref{theory}. In the rest of the paper, electron
Volts (eV) and femtoseconds (fs) are used for energy and time, respectively, unless otherwise specified.

\section{Theory}\label{theory}

\subsection{General description of the model  system}\label{general}

Consider an atomic system with ground state $|g\rangle$ and two doubly excited states $|a\rangle$ and $|b\rangle$. The dipole
transition is allowed between $|g\rangle$ and $|a\rangle$ and between $|a\rangle$ and $|b\rangle$. Suppose their energy levels and
the external fields are arranged in a way such that the XUV is near resonance between $|g\rangle$ and $|a\rangle$ and the laser is
near resonance between $|a\rangle$ and $|b\rangle$. The total time-dependent wave function of the system can be approximately
written as
\begin{align}
|\Psi(t)\rangle &= e^{-i E_g t} c_g(t) |g\rangle \notag\\
&+ e^{-i E_X t} \left[ d_a(t) |a\rangle + \int{d_{E_1}(t) |E_1\rangle dE_1} \right] \notag\\
&+ e^{-i E_L t} \left[ d_b(t) |b\rangle + \int{d_{E_2}(t) |E_2\rangle dE_2} \right], \label{total}
\end{align}
where $E_g$ is the ground state energy, $E_X \equiv E_g+\omega_X$ and $E_L \equiv E_g+\omega_X+\omega_L$ are the ``pumped energies''
with respect to the photon energies $\omega_X$ and $\omega_L$, where $X$ is for XUV and $L$ is for laser; $|E_1\rangle$ and
$|E_2\rangle$ are the continua with respect to $|a\rangle$ and to $|b\rangle$. The fast oscillating parts in the wave function have
been factored out, and the $c(t)$ and $d(t)$ functions are assumed slowly-varying with time. By convention, we use symbol
$c(t)$ for coefficients if the expansion is with respect to eigenstates, and symbol $d(t)$ is used when the expansion is in terms of
configurations (not eigenstates). The total Hamiltonian of the atomic system in the external fields is
\begin{equation}
H(t) = H_A + H_X(t) + H_L(t), \label{ham}
\end{equation}
where $H_A$ is the atomic Hamiltonian, and $H_X(t)$ and $H_L(t)$ represent the interactions of the XUV and laser fields with the
atom, respectively. In the photon energy range in consideration, the interaction with the field is given by the electric dipole
transition. Equation~\ref{ham} provides the total Hamiltonian to solve the time-dependent Schr\"{o}dinger equation for the wave
function in Eq.~\ref{total}, i.e.,
\begin{equation}
i\frac{d}{dt}|\Psi(t)\rangle = H(t)|\Psi(t)\rangle. \label{tdse}
\end{equation}
The fields are assumed linearly polarized in the same direction. The
electric fields are in the form of
\begin{align}
E(t) &= 2F(t) \cos (\omega t) \notag\\
&= F(t) \left( e^{i \omega t} + e^{-i \omega t} \right), \label{field}
\end{align}
where $2F(t)$ is the pulse envelope. The present model works for arbitrary pulse envelopes; however, we limit
the envelope to cosine-square shape in the calculation, i.e.,
\begin{equation}
F(t) = F_0 \cos^2 \left( \frac{t-t_0}{\tau} \right) \;\text{for}\; -\frac{\pi \tau}{2} < t-t_0 < \frac{\pi \tau}{2}
\label{envelope}
\end{equation}
and $F(t)=0$ anywhere else, where $t_0$ is the peak time. For such a pulse, the pulse duration is $\pi \tau /2.75$, and $F_0$ is
related to the peak intensity $I_0$ by $I_0=4F_0^2$. In this paper, the pulse duration is defined by the full width at half maximum
(FWHM) of the intensity envelope. The time delay between the two pulses is define as the time of laser peak subtracted by the time
of XUV peak, so it is positive if the peak of the  XUV appears before the IR. By convention, in this paper, the XUV peak and the IR
peak are placed at $t=0$ and $t=t_0$ respectively, so the time delay is $t_0$.

\begin{figure}
\centering
\includegraphics[width=0.35\textwidth]{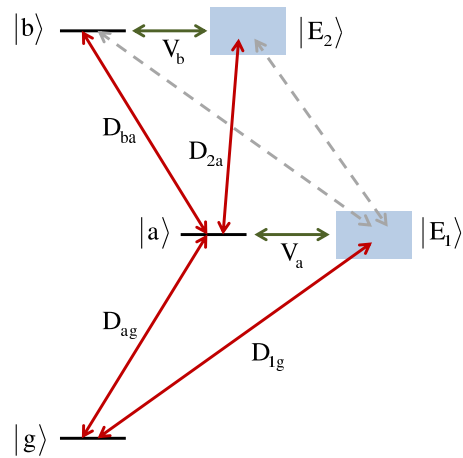}
\caption{(Color online) Diagram for off-diagonal terms of the Hamiltonian in the configuration basis. The double-arrows represent the
transitions between the basis vectors; the red ones represent the dipole transitions, the green ones represent the CIs responsible for
autoionization, and the gray dashed ones are neglected in the model.}
\label{fig-sys}
\end{figure}

Note that the basis functions for the autoionizing states in Eq.~\ref{total} are not the eigenstates of the atomic Hamiltonian
$H_A$. In the basis space of Eq.~\ref{total}, the diagonal terms of the Hamiltonian are just the energies $E_g$, $E_a$, $E_1$, $E_b$,
and $E_2$. The off-diagonal terms are
\begin{align}
\langle E_1 |H(t)| a \rangle &= V_a \label{ham1} \\
\langle E_2 |H(t)| b \rangle &= V_b, \\
\langle a |H(t)| g \rangle &= -D_{ag} F_X(t) e^{-i \omega_X t} \\
\langle E_1 |H(t)| g \rangle &= -D_{1g} F_X(t) e^{-i \omega_X t} \\
\langle b |H(t)| a \rangle &= -D_{ba} F_L(t) e^{-i \omega_L t} \\
\langle E_2 |H(t)| a \rangle &= -D_{2a} F_L(t) e^{-i \omega_L t} \label{ham2}
\end{align}
where the $D$ are the dipole matrix elements and $V$  the transition amplitudes of configuration interaction (CI). These terms are
schematically plotted in Fig.~\ref{fig-sys}.

For the process in concern, additional simplifications are made. Helium atom is taken as a prototype, where $|g\rangle$ is
$1s^2(^1S)$, $|a\rangle$ and $|b\rangle$ are $2s2p(^1P)$ and $2p^2(^1S)$, and $|E_1\rangle$ and $|E_2\rangle$ are $1s\epsilon p(^1P)$
and $1s\epsilon s(^1S)$, respectively. Referring to Fig.~\ref{fig-sys}, since the IR coupling is a one-electron dipole operator, the
coupling between $2p^2(^1S)$ and $1s\epsilon p(^1P)$ is zero to the first order. The coupling between the two continuum states by the
dipole operator can also be neglected since the IR is not absorbed by the continuum electron. We also use the rotating wave
approximation such that only the resonant transitions are considered. Furthermore, the matrix elements involving continuum states
$|E_1\rangle$ and $|E_2\rangle$ are assumed energy-independent, which means $D$ and the Fano parameters, $\Gamma$ and $q$, are
constant values estimated at the resonances. This is a good approximation when the resonance energy is high enough above the
threshold relative to its width so that the continuum only varies slightly across the resonance. In Eqs.~\ref{ham1}-\ref{ham2}, the
basis functions are real by convention (the continua are standing waves) so that all the $D$ and $V$ are real. In this model, all the
atomic parameters ($D$, $E$, $\Gamma$, and $q$) are taken from experiments or from calculations in the literature. These parameters
are determined by the atomic structure which is irrelevant to the development of our model.

\subsection{Three-state model}\label{bound}

Under the approximations outlined above, a system of coupled equations for all the coefficients appearing in Eq.~\ref{total} are obtained. Considering the conservation of
the total probability of the wave packet, the continuum-state coefficients change with time much more slowly than the bound-state coefficients. Thus, $d_{E_1}(t)$ and
$d_{E_2}(t)$ are adiabatically eliminated by assuming their time-derivatives are zeros~\cite{madsen,themelis}. This allows us to reduce the calculation to include only
the bound states. Solving the coupled equations is then numerically feasible. Now we have
\begin{align}
i\dot{c}_g(t) &= -i\frac{\gamma_g(t)}{2} c_g(t) + \lambda_a F_X(t) d_a(t) \label{c_g} \\
i\dot{d}_a(t) &= \lambda_a F_X(t) c_g(t) - \left[\delta_X + i\frac{\Gamma_a+\gamma_a(t)}{2}\right]
d_a(t) \notag\\
&+ \lambda_b F_L(t) d_b(t) \label{d_a} \\
i\dot{d}_b(t) &= \lambda_b F_L(t) d_b(t) - \left(\delta_X + \delta_L + i\frac{\Gamma_b}{2}\right) d_b(t), \label{d_b}
\end{align}
where $\lambda_a \equiv -D_{ag} (1-i/q_a)$ and $\lambda_b \equiv -D_{ba} (1-i/q_b)$ are the newly defined complex dipole matrix
elements which include the route through the continua. The laser-induced broadening $\gamma_g(t)$ and $\gamma_a(t)$ are defined by
\begin{align}
\gamma_g(t) &\equiv 2\pi \left| D_{1g} F_X(t) \right|^2 \\
\gamma_a(t) &\equiv 2\pi \left| D_{2a} F_L(t) \right|^2,
\end{align}
the detuning of the pulses are $\delta_X \equiv \omega_X + E_g - E_a$ and $\delta_L \equiv \omega_L + E_a - E_b$,
and the Fano parameters are $\Gamma_a \equiv 2\pi |V_a|^2$, $\Gamma_b \equiv 2\pi |V_b|^2$,
$q_a \equiv D_{ag}/(\pi V_a D_{1g})$, and $q_b \equiv D_{ba}/(\pi V_b D_{2a})$. The AC Stark shifts vanish, i.e.,
\begin{equation}
P \int{\frac{|D_{1g}F_X(t)|^2}{E_X-E_1}dE_1} = P \int{\frac{|D_{2a}F_L(t)|^2}{E_L-E_2}dE_2} = 0
\end{equation}
because $D_{1g}$ and $D_{2a}$ are constant of energy, where $P$ means the principal  part.

With any given set of atomic parameters, field parameters, and initial conditions, Eqs.~\ref{c_g}-\ref{d_b} uniquely
determine the bound-state part of the total wave function regardless of the continua. In numerical calculations, we
propagate these coefficients using the Runge-Kutta method, with the initial conditions $c_g(t_i)=1$ and
$d_a(t_i)=d_b(t_i)=0$, i.e., the system is in the ground state at initial time $t_i$ before the pulses arrive.
Note that the model described above retrieves only the bound-state part of the total wave function. It has been applied successfully
for long pulses where the photoelectron distribution in each single measurement is disregarded~\cite{bachau, madsen, themelis, loh}.
In the present study, however, a broadband XUV pumps the electrons to the continuum and to the bound state alike; missing the
continuum coefficients means missing the essential information of the total wave function. In the following, we aim to extend the
model to recover the dynamics of the complete system for short-pulse cases.

\subsection{Continuum states }\label{continuum}

The original coupled equations for the continuum states before the adiabatic elimination are
\begin{align}
i\dot{d}_{E_1}(t) &= -D_{1g} F_X(t) c_g(t) + V_a d_a(t) + (E_1-E_X) d_{E_1}(t) \label{d_E1} \\
i\dot{d}_{E_2}(t) &= -D_{2a} F_L(t) d_a(t) + V_b d_b(t) + (E_2-E_L) d_{E_2}(t) \label{d_E2}.
\end{align}
Assuming $\dot{d}_{E_1}(t)=\dot{d}_{E_2}(t)=0$ in Eqs.~\ref{d_E1}-\ref{d_E2}, the approximate $d_{E_1}(t)$ and $d_{E_2}(t)$ have
singularities at $E_1=E_X$ and at $E_2=E_L$, respectively. When plugged into the bound-state coupled equations, these singularities
are handled properly with contour integrations. After solving Eq.~\ref{c_g}-\ref{d_b}, the bound-state coefficients $c_g(t)$,
$d_a(t)$, and $d_b(t)$ are known functions of time. We then return to the full version of the continua in Eq.~\ref{d_E1}-\ref{d_E2}
to retrieve $d_{E_1}(t)$ and $d_{E_2}(t)$, again by the Runge-Kutta propagation over time, with the initial conditions
$d_{E_1}(t_i)=d_{E_2}(t_i)=0$. The new $d_{E_1}(t)$ and $d_{E_2}(t)$ functions generated by Eq.~\ref{d_E1}-\ref{d_E2} are
the next iteration and better solutions to the previous ones.

After the pulses are over, the system may still change due to autoionization. For an arbitrary set of bound- and continuum-state
coefficients, the photoelectrons evolve to a final energy distribution after about 10 times the decay lifetime or so~\cite{chu}. To
simulate this measurable photoelectron spectrum, we have to project the total wave packet that has propagated for a long time onto
the partial waves of interest; equivalently, taking the $|a\rangle$-$|E_1\rangle$ resonance for example, the spectrum is
proportional to $|d_{E_1}(t_f)|^2$ as $t_f \to \infty$. The probability density of the continuum wave function, $P(E_1) \equiv
|d_{E_1}(t_f)|^2$, is called ``photoelectron profile'' or simply ``profile'' hereafter in this paper. Numerically, $t_f$ is taken
when $|d_{E_1}(t)|^2$ does not change anymore. In the case concerning the $2s2p(^1P)$ resonance in helium, the decay lifetime
is 17~fs, so the final spectrum can be safely taken at about $t_f=200$~fs. Note that $P(E_1)$ has the dimension of probability
per unit of energy, and the integral of $P(E_1)$ over energy represents the total probability of the system being in the
$|E_1\rangle$ continuum. The conservation of total probability can be used to numerically check the convergence of calculation. The
continuum states calculated this way have some advantages. First, each of them can be turned on separately once the bound-state
calculation is done. Second, the energy range and energy mesh can be arbitrarily chosen without lowering the accuracy because
$d_{E_1}(t)$ and $d_{E_2}(t)$ are calculated independently for each energy point.

In the present work, we only concern ourselves with the final photoelectrons reaching the detector. However, if one wants to know
the evolution of electrons in real time before the end of decay, the retrieval of $|d_{E_1}(t)|^2$ and $|d_{E_2}(t)|^2$ as functions
of energy and time will be the answer, instead of $P(E_1)$ and $P(E_2)$ as just functions of energy. The measurement of this
short-time behavior was proposed using an additional high-energy short pulse to ionize the inner electron~\cite{chu}.

\subsection{Eigenstates}\label{eigen}

The atomic eigenstates near a resonance are solved in terms of the corresponding bound state and background continuum by Fano's
theory~\cite{fano}. This leads us to express the total wave function in eigenstate basis in the general form of
\begin{align}
|\Psi(t)\rangle &= e^{-i E_g t} c_g(t) |g\rangle + e^{-i E_X t}  \int{ c_E^{(a)}(t) | \psi_E^{(a)} \rangle dE } \notag\\
&+ e^{-i E_L t}  \int{ c_{E'}^{(b)}(t) | \psi_{E'}^{(b)} \rangle dE' }, \label{total-eig}
\end{align}
where the superscripts $(a)$ and $(b)$ indicate the $|a\rangle$-$|E_1\rangle$ pair and the $|b\rangle$-$|E_2\rangle$ pair,
respectively. The eigenstates we refer to are the eigenstates of atomic Hamiltonian $H_A$, so they are stationary only in the
absence of field. In the following, we will discuss only the subspace spanned by $|a\rangle$ and $|E_1\rangle$ in details, while
the same principles apply to $|b\rangle$ and $|E_2\rangle$. In the meantime the superscript $(a)$ is omitted.

The solutions for an eigenstate of energy $E$ in configuration basis is
\begin{equation}
|\psi_E\rangle = \mu_E |a\rangle + \int{\nu_{EE_1} |E_1\rangle dE_1}, \label{eig-config}
\end{equation}
where the coefficients are
\begin{align}
\mu_E &= \frac{\sin \theta_E}{\pi V_a} \\
\nu_{EE_1} &= \frac{\sin \theta_E}{\pi (E-E_1)} - \cos \theta_E \delta (E-E_1),
\end{align}
where
\begin{equation}
\theta_E \equiv -\tan^{-1} \left( \frac{\Gamma_a /2}{E-E_a} \right).
\end{equation}
The conversion between Eq.~\ref{total} and Eq.~\ref{total-eig} means that for the $|a\rangle$ resonance,
\begin{equation}
\int{c_E(t) |\psi_E\rangle dE} = d_a(t) |a\rangle + \int{d_{E_1}(t) |E_1\rangle dE_1}. \label{replace}
\end{equation}
Starting with Eq.~\ref{eig-config} and Eq.~\ref{replace}, with some algebra, the $c_E(t)$ coefficients are
\begin{equation}
c_E(t) = \frac{\sin \theta_E}{\pi V_a} d_a(t)-(\cos \theta_E-i\sin \theta_E) d_{E_1}(t)\big|_{E_1=E}, \label{c_E}
\end{equation}
which enables us to calculate the detected photoelectron spectrum $P(E) \equiv |c_E(t_f)|^2$ as $t_f \to \infty$. This spectrum
defined here will reach the same result as $P(E_1)$ defined in Sec.~\ref{continuum}, but just from a different calculation
procedure.

The profile $|c_E(t)|^2$ evolves differently from $|d_{E_1}(t)|^2$. As pointed out in Sec.~\ref{continuum}, $|d_{E_1}(t)|^2$ evolves
until the end of the decay, and $|c_E(t)|^2$ evolves until the field is over. In a typical case where the field ends much earlier
than the end of the decay, it is  more efficient to calculate the final $|c_E(t_f)|^2$ than  the final $|d_{E_1}(t_f)|^2$. Moreover,
in Eq.~\ref{c_E}, for a point in time and in energy $(t,E)$, the coefficient $c_E(t)$ is calculated by a simple algebra with the
$d_a(t)$ and $d_{E_1}(t)|_{E_1=E}$ of just the same $(t,E)$; there is no integral over energy nor propagation over time to carry out
$c_E(t)$ once $d_a(t)$ and $d_{E_1}(t)|_{E_1=E}$ are known. In other words, including $c_E(t)$ in the calculation requires very
little extra efforts. Summing up these facts, for the purpose of retrieving the electron spectra, it is advantageous to adopt
$P(E)=|c_E(t_f)|^2$ instead of keeping the form of $P(E_1)=|d_{E_1}(t_f)|^2$. We have come to the conclusion that for
the present system and setup, the optimal workflow is in the consecutive order of bound states, continuum states, and
eigenstates, all of which calculated in the physical time until the field vanishes.

For the subspace spanned by $|b\rangle$ and $|E_2\rangle$, the eigenstate coefficients are constructed by $d_b(t)$ and $d_{E_2}(t)$
in the same way but with different parameters. Now both $c_E^{(a)}(t)$ and $c_E^{(b)}(t)$ are obtained, and we can fully build the
total wave function in eigenstates basis as shown in Eq.~\ref{total-eig}.

\section{Results and discussions}\label{results}

\subsection{Laser wavelength of 780~nm}\label{780nm}

\subsubsection{Calculation and analysis}\label{780nm_cal}

We first apply the model to the case where the laser couples $2s2p(^1P)$ and $2p^2(^1S)$ in helium with the experimental setup
reported by Gilbertson \textit{et al}~\cite{gilbertson}. The two doubly excited states autoionize to $1s\epsilon p(^1P)$ and
$1s\epsilon s(^1S)$ respectively. The measured quantity is the electron spectrum, corresponding to $P(E)=|c_E(t_f)|^2$ in the model
introduced in Sec.~\ref{eigen}, where $t_f$ is taken after the field vanishes. For the XUV pulse, the photon energy is 60~eV and the
duration is 100~as; the bandwidth is 20~eV, which is much wider than the width of $2s2p$, and viewed as a flat background in spectrum
near the resonance. The XUV transition from the $1s^2(^1S)$ ground state to $2s2p$ is nearly resonant, and we conveniently set
$\delta_X=0$. In principle, such pump pulse can initiate other $2snp$, $2pns$ and $2pnd$ states which should all be included in the
total wave function. (For more accurate description of doubly excited states, see Lin~\cite{lin86}) Nonetheless, those higher
states have longer lifetimes, exhibiting less dynamics in the time scale in our scheme; furthermore, their resonance widths are
narrow and difficult to measure with typical electron spectrometers. Thus, we treat $2s2p$ as the only $|a\rangle$ state in
Eq.~\ref{total}, while $1s\epsilon p$ is $|E_1\rangle$. The Fano parameters of the $2s2p$ resonance are experimental values taken
from the literature~\cite{domke96}, where $E_a=60.15$~eV, $\Gamma_a=37$~meV, and $q_a=-2.75$. The XUV is weak and in the linear
regime so that its intensity does not affect the spectra other than an overall factor. The peak XUV intensity is assumed
$I_X=10^{10}$~W/cm$^2$.

The laser pulse is a 9 fs IR pulse with wavelength $\lambda_L=780$~nm and peak intensity $I_L=7\times10^{11}$~W/cm$^2$. With
1.6~eV photon energy, the IR field couples $2s2p$ and $2p^2$ at 62.06~eV, where $\delta_L=-0.4$~eV. The atomic parameters
$\Gamma_b=5.9$~meV~\cite{burgers} and $D_{ba}=2.17$~a.u.~\cite{loh} are taken from literature. The bound-free dipole matrix element
$D_{2a}$ responsible for the transition from $2s2p$ to $1s\epsilon s$ is very small because it is a second order (satellite)
transition and requires electron correlation, while $D_{ba}$ between $2s2p$ and $2p^2$ is a first-order transition via $\langle
2s|D|2p \rangle$. The parameter $q_b$, representing the ratio of $D_{ba}$ to $D_{2a}$, is thus very large; it is set $q_b=1000$.
Note that when $|q| \gg 1$, the Fano shape appears as a symmetric peak, and the sign of $q$ is insignificant.

In order to understand the dynamics controlled by the laser, we evaluate the generalized Rabi frequency defined by
\begin{equation}
\Omega(t) \equiv \sqrt{|D \cdot E(t)|^2 + |\delta|^2}. \label{rabi}
\end{equation}
When the detuning is large, the Rabi frequency is higher; however, the amplitude of the oscillation is lower, i.e., the population
does not fully swing to the other coupled state. In our system and setup, $\Omega=0.43$~eV between $2s2p$ and $2p^2$ at the laser
peak, corresponding to 9.6~fs period. For the two doubly excited states, the binding energies of $2s2p$ and $2p^2$ are 5.3 and
3.3~eV, which are low enough such that their ionizations are quick, especially the latter one. We calculate the ionization rates for
both states using the model developed by Perelomov, Popov, and Terent'ev~\cite{ppt}, referred by the PPT model hereafter, with the
high-intensity correction introduced in~\cite{tong05}. The empirical parameters in our
PPT calculation are obtained by fitting our result to the experiment, which will be discussed in Sec.~\ref{780nm_exp}. The
calculated peak rates for $2s2p$ and $2p^2$, in units of energies, are 5.4~meV and 0.46~eV, comparable to their resonance widths
37~meV and 5.9~meV, respectively. To incorporate these ionization rates to our coupling model in Eqs.~\ref{c_g}-\ref{d_b}, the
widths are broadened by $\Gamma_{a,b}'(t)=\Gamma_{a,b}+W_{a,b}(t)$ where $W_a(t)$ and $W_b(t)$ are the time-dependent ionization
rates for $2s2p$ and $2p^2$ respectively. Note that for the dynamics we have considered so far, on one hand, the coupling and the
ionization of resonances exist only in the presence of laser field; on the other hand, the autoionization processes are determined
exclusively by the atomic structure, whose time scale can not be changed externally.

\begin{figure}
\centering
\includegraphics[width=0.45\textwidth]{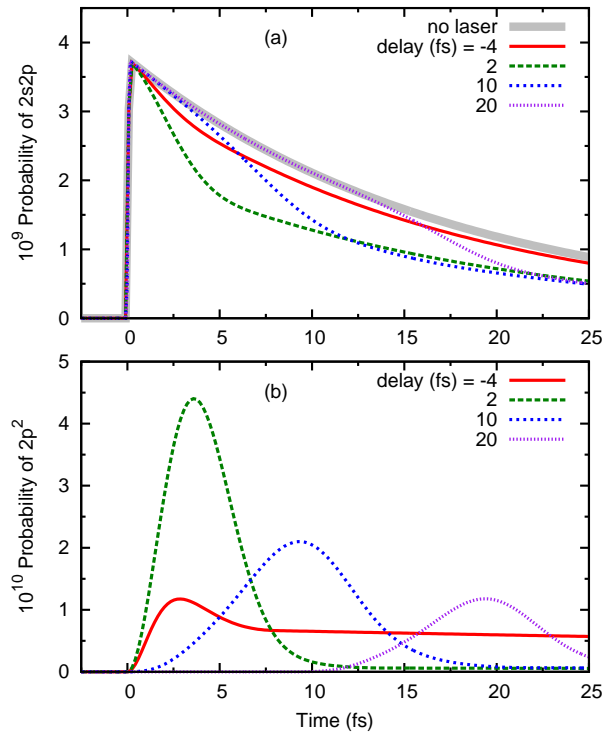}
\caption{(Color online) Probabilities of the (a) $2s2p$ and (b) $2p^2$ bound states with time, for $\lambda_L=780$~nm and various
$t_0$. The case without laser is plotted in the gray solid curve in (a).}
\label{fig-case1-bound}
\end{figure}

The probabilities of the $2s2p$ and $2p^2$ bound states are presented in Fig.~\ref{fig-case1-bound} as they propagate in time.
Although the bound states are not directly measured in this pump-probe scheme, the analysis therein helps reveal the physics behind
the whole time-dependent process. We will analyze the bound-state propagation and the resonance profiles in the following at the
same time.

\begin{figure}
\centering
\includegraphics[width=0.45\textwidth]{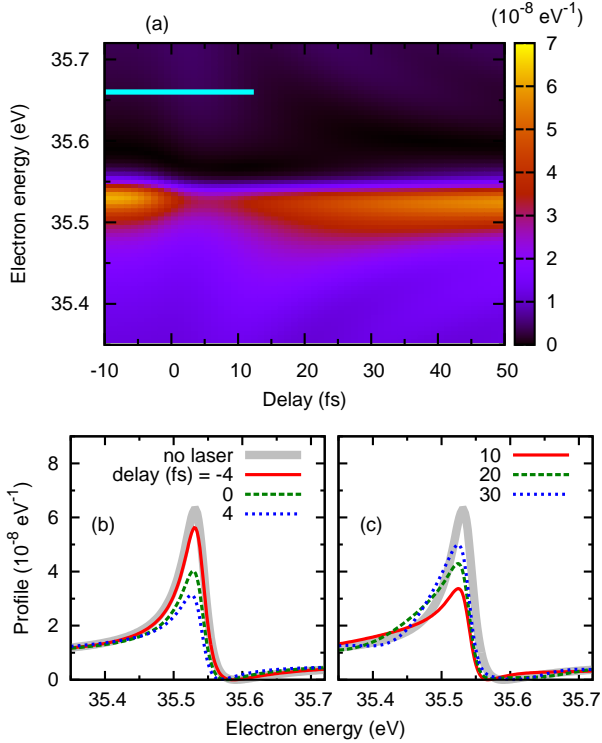}
\caption{(Color online) Photoelectron profile of the $2s2p$ resonance for $\lambda_L=780$~nm. (a) The spectrogram. The cyan bar
indicates the range of delay where the two pulses overlap. (b) The spectra for various $t_0$, and the case without laser.}
\label{fig-case1-a}
\end{figure}

The photoelectron profiles of the $2s2p$ resonance for $t_0=-10$ to 50~fs are shown in Fig.~\ref{fig-case1-a}. If the laser field is
absent, the dynamics of the system after the pump will be nothing more than the autoionization of $2s2p$, and the detector will
see an original Fano line-shape as seen in the absorption spectrum in synchrotron radiation experiment. When a laser is added very
early, e.g., $t_0=-10$~fs, the spectrum only changes insubstantially because the laser is already gone by the time the XUV pumps the
system. The laser has essentially no influence on the dynamics; this is seem at the left end of the spectrogram. If the laser is
shifted after the XUV, the laser will start pumping the system from $2s2p$ to $2p^2$. The period of Rabi oscillation is 9.6~fs, close
to the laser duration 9~fs. The ionization of $2p^2$ has the time scale $1/W_b=1.4$~fs estimated at the laser peak. Such a short time
indicates that $2p^2$ is very quickly depleted in the presence of the laser. For $t_0=0$ to 5~fs, the laser is mainly at the
beginning of decay. Most population is brought from $2s2p$ to $2p^2$ before $2s2p$ decays, and being ionized from $2p^2$ without returning to $2s2p$. Figure~\ref{fig-case1-bound}(a) shows that for $t_0=2$~fs, there is a quick drop at around $t=4$~fs which takes
away about 40\% of the original population if compared to the ``no laser'' level. The amount of the $2s2p$ doubly excited state is
greatly reduced, generating much less photoelectrons and creating a significantly lower peak in the spectrum, as seen in
Fig.~\ref{fig-case1-a}(a). Figure~\ref{fig-case1-a}(b) shows the change of the resonance profile vs the time delays in this
overlapping region. If the laser moves further positively, the longer lags between the two pulses gives more time for $2s2p$ to decay
before the coupling kicks in. For $t_0 \gg 17$~fs, the laser is late enough so that $2s2p$ decays completely without being
interrupted, i.e., the original Fano shape is fully restored, and the laser has nothing to pump anymore. The recovery of the
resonance profile vs time delay is shown in Fig.~\ref{fig-case1-a}(c). Ultimately, because the laser only removes bound electrons,
changing the delay time traces out the decay process.

\begin{figure}
\centering
\includegraphics[width=0.45\textwidth]{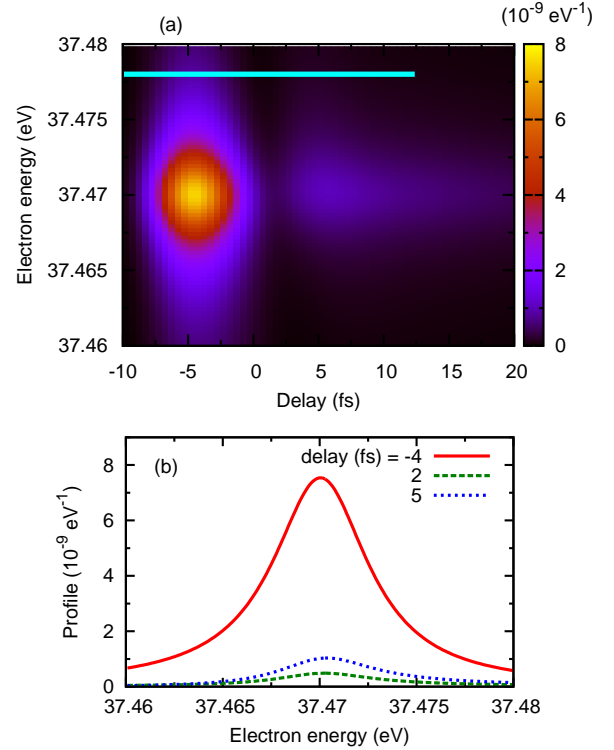}
\caption{(Color online) As Fig.~\ref{fig-case1-a} but for the $2p^2$ resonance.}
\label{fig-case1-b}
\end{figure}

Figure~\ref{fig-case1-b} shows the $2p^2$ resonance profiles. The signal is large in the range from $t_0=-10$~fs to 0 and low for
$t_0>15$~fs. Starting from the left end of the spectrogram, for $t_0=-4$~fs, the tail part of the laser pulse is involved in the
dynamics; counting only this involved part of the laser, the strength is low and the duration is short, and the Rabi oscillation
is slow. As a result, the maximum population pumped to $2p^2$ is ``trapped'' without returning back to $2s2p$. At the same time, the
ionization rate by this laser ``tail'' is low and not influential. The high population of the $2p^2$ state is shown in
Fig.~\ref{fig-case1-bound}(b). At large time, it decays into the biggest photoelectron profile shown in Fig.~\ref{fig-case1-b}(b).
Moving on to $t_0=2$~fs, the laser strikes $2s2p$ mainly at the beginning of decay. As shown in Fig.~\ref{fig-case1-bound}(b), the
full strength of laser depletes the population almost completely; at the end of laser, the $2p^2$ bound state is almost empty, and
there are hardly any electrons to autoionize. Consequently, the $2p^2$ profile in Fig.~\ref{fig-case1-b} is greatly depressed,
forming a valley in the spectrogram.

Note that the $2s2p$ and the $2p^2$ resonances are in different symmetries.  Their momentum (or angular) distributions are
different, but their energy ranges overlap because of the broad bandwidth of the SAP. If the momentum spectrum is measured, one can
distinguish the contributions from the two resonances and separate out the spectra. However, if only the energy spectrum is measured,
the two contributions are not separable near 37.5~eV, where the $2s2p$ resonance lies on the $1s\epsilon p$ background. The
$1s\epsilon p$ background at 37.5~eV is $7.5 \times 10^{-9}$~eV$^{-1}$, which is about the same as the peak value of the $2p^2$
profile shown in Fig.~\ref{fig-case1-b}. The $2p^2$ resonance profile in the total energy spectrum will appear flatter than how it
would look without the background.

\begin{figure}
\centering
\includegraphics[width=0.45\textwidth]{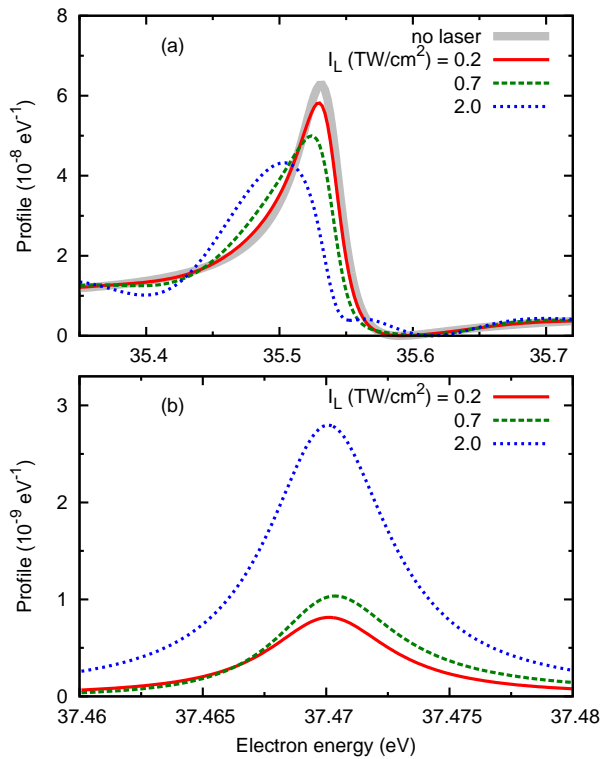}
\caption{(Color online) Photoelectron profiles for $\lambda_L=780$~nm and various $I_L$, of (a) the $2s2p$ resonance for $t_0=30$~fs,
and (b) the $2p^2$ resonance for $t_0=5$~fs.}
\label{fig-case1-I}
\end{figure}

The mechanism of the dressing laser is further analyzed  when the calculations are done for various laser intensities. The curves in
Fig.~\ref{fig-case1-I}(a) and (b) represent the spectra for fixed delays but different intensities, where the medium one
($I_L=0.7$~TW/cm$^2$) repeats the experimental condition. The delays 30 and 5~fs are chosen for $2s2p$ and $2p^2$ respectively for
adequate signal strengths. For $2s2p$, with the increase of $I_L$, the profile gradually depresses and forms interference patterns. It
suggests that if the laser at about 30~fs is strong enough, it will suddenly deplete the bound states and halt the decay process. The
resultant profile is then constructed by the electrons autoionized before 30~fs. Fig.~\ref{fig-case1-I}(a) also shows that
the resonance peak moves gradually to the low-energy side with increasing IR coupling strength. For the $2p^2$ resonance shown in
Fig.~\ref{fig-case1-I}(b), increasing $I_L$ mainly populates more $2p^2$ and makes the profile larger.

\subsubsection{Comparison with experiment}\label{780nm_exp}

\begin{figure}
\centering
\includegraphics[width=0.45\textwidth]{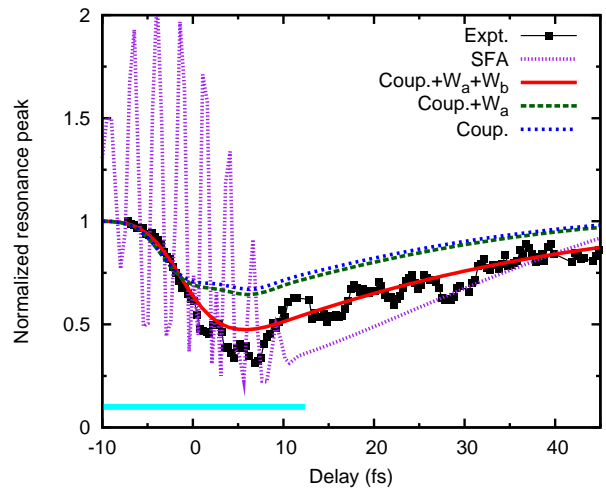}
\caption{(Color online) The normalized peak value of the $2s2p$ profile as a function of delay for $\lambda_L=780$~nm. The value for
$t_0=-10$~fs is
set as 1. The present calculations are shown in three different settings where the ionization rates $W_a(t)$ and $W_b(t)$ are
included or excluded. The experimental curve~\cite{gilbertson} is shifted in $t_0$ by 4.5~fs to fit the full calculation shown by
the red solid curve. The result by the SFA model~\cite{zhao} is normalized such that it fits the experimental data the most. The
cyan bar indicates the delay range where the two pulses overlap.}
\label{fig-exp-dl}
\end{figure}

The experiment by Gilbertson \textit{et al}~\cite{gilbertson} reported the spectrogram in the energy range 33--46~eV, enclosing both
resonances in our concern. However, the energy resolution is insufficient for the very narrow $2p^2$ resonance with
$\Gamma_b=5.9$~meV, making its transient spectra invisible. The most dominant feature in the spectrogram is the variation in peak
height of $2s2p$ resonance, where the decay lifetime 17~fs can be extracted. In Fig.~\ref{fig-exp-dl}, the normalized peak value is
shown as a function of $t_0$. The peak value for the most negative delay, which represents the ``XUV only'' case, is normalized to 1.
The present calculation is displayed in three different settings, where the model excludes the laser ionization, includes the laser
ionization from $2s2p$, and includes the laser ionization from both states. While the experiment did not know the absolute times of
the pulses thus unable to determine the absolute delay, the experimental data is shifted by 4.5~fs to fit our calculation. In
Fig.~\ref{fig-exp-dl}, the calculation result without laser ionization exhibits the similar decay feature found in the
experiment, but the absolute value is too high. By adding the laser ionization, we are able to adjust the parameters such that the
calculation agrees with the experiment quantitatively. The empirical ionization parameters are $C_l=0.5$ and $\alpha=0$ for $2s2p$ and
$C_l=0.5$ and $\alpha=11$ for $2p^2$, where $C_l$ is defined in the PPT model~\cite{ppt} and $\alpha$ is defined in the correction
term~\cite{tong05}.

The comparison shown above indicates that the presence of the $2p^2$ state is responsible for the significant depression of the Fano
peak, where both the coupling and the IR ionization take credit. The agreement between our model
and the experiment is very good when both effects are taken into account. However, there are irregular oscillations in the
experimental data; their period is typically between 5 and 10~fs, but mostly unpredictable. Their appearance is unexplained so far,
but could be due to experimental artifact.

\begin{figure}
\centering
\includegraphics[width=0.45\textwidth,clip,trim=0 0 0 5mm]{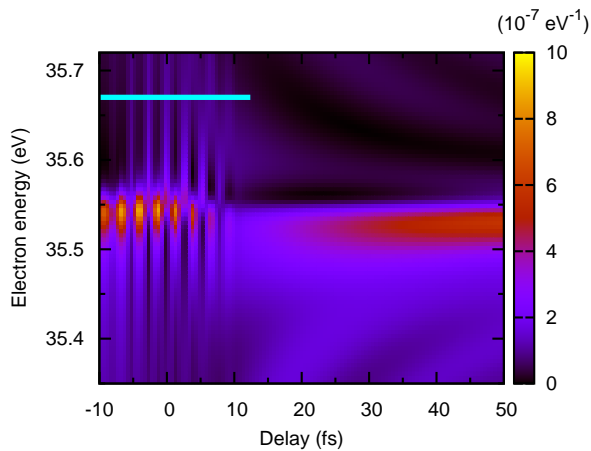}
\caption{(Color online) The angular-differential spectrogram of the $2s2p$ resonance in the direction of the laser polarization
calculated by the SFA model~\cite{zhao}.}
\label{fig-sfa}
\end{figure}

The experimental report included a simulation using the "streaking" model (within the SFA) for isolated autoionizing
states~\cite{zhao}, with the inclusion of ionization due to the laser calculated within the PPT model~\cite{ppt}, as described by
Eq.~1 in Ref.~\cite{gilbertson}. In the SFA model, the ionization of $2s2p$ and the acceleration of the scattered electrons
are treated separately, and the laser coupling to other bound or resonance states is totally disregarded. The common mechanism
between the model therein and our model is that the laser ionization is treated by the PPT calculation. In order to have a clear view
on the two approaches, we have reproduced the angular-differential electron spectrogram at the polarization direction, where the
measurement was done, using the SFA model with the same parameters taken in our model. The result is shown in Fig.~\ref{fig-sfa}. The
corresponding resonance peak is plotted in Fig.~\ref{fig-exp-dl} which is normalized to fit the overall experimental data. The SFA
results in Fig.~\ref{fig-exp-dl} and in Fig.~\ref{fig-sfa} are similar to our calculation for $t_0>10$~fs; the resonance peak goes
through the same depression before the gradual revival along $t_0$. However, for $t_0<10$~fs, the strong streaking peaks carrying the
laser period 2.6~fs are missing both in the experiment and in our model. This suggests that with the inclusion of coupling and
ionization of $2p^2$, the present model takes charge of the primary effects seen in the time-delayed measurement. We comment that even
though both the SFA model and the present one account for the ionization by the IR laser, it is the ionization of $2p^2$ that is
mostly responsible for the great reduction in the resonance strength observed in the experiment. This state was not included in the
SFA model.

\begin{figure}
\centering
\includegraphics[width=0.45\textwidth]{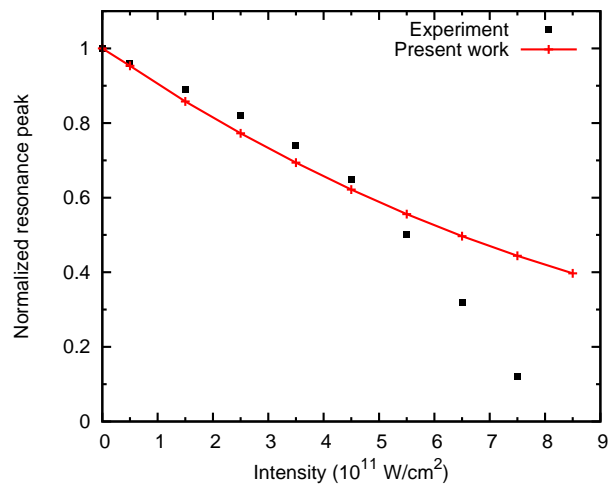}
\caption{(Color online) The normalized peak value of the $2s2p$ profile at the dip (see text) as a function of the peak intensity of
laser.}
\label{fig-exp-I}
\end{figure}

The experiment also measured the spectrograms with several other laser intensities, and reported the depth of the ``dip'' that is
seen in Fig.~\ref{fig-exp-dl}'s curve against $I_L$. The experimental and theoretical dip values are shown in Fig.~\ref{fig-exp-I}.
As seen in the figure, increasing $I_L$ will aggravate the depression of the resonance. The experimental dip is also deeper than the
theoretical one for high $I_L$, i.e., the resonance peak is more depleted than what the present theory predicts. This is because
as $I_L$ increases, the nonlinear interaction of the IR opens up more excitation and ionization pathways that deplete the $2s2p$ part
(or the bound part) of the resonance. The depletion will lead to fewer autoionization events and thus weaker electron signal near the
resonance. Note that the discrepancy between the present model and experimental data in Fig.~8 grows nonlinearly with $I_L$
for $I_L > 4.5 \times 10^{11}$~W/cm$^2$.

\subsection{Laser wavelength of 540~nm}\label{540nm}

Using long pulses and laser spectroscopy, Autler-Townes doublet~\cite{autler,fleischhauer} has been observed when two bound states
are strongly coupled by a dressing laser field. For the two autoionizing states and the ultrashort laser in the present study,
can we observe anything resembling the Autler-Townes doublet in the spectra? In Sec.~\ref{780nm}, we have shown that for
the 780~nm IR, since the laser detuning is large and the laser ionization is significant, the evidence of Autler-Townes doublets is
not prevalent except that the small shift of the $2s2p$ peak at higher coupling intensity in Fig.~\ref{fig-case1-I}(a) gives a feeble
hint. Now, we intentionally tune the laser to $\lambda_L=540$~nm (mean photon energy is 2.30~eV) so that it is in resonance between
$2s2p$ and the lower-energy state, $2s^2(^1S)$ at $57.85$~eV. In this setup, first of all, the laser detuning is negligible, making
the Rabi oscillation the strongest. Secondly, the binding energy of the $|b\rangle$ state changes from 3.34~eV for $2p^2$ to 7.55~eV
for $2s^2$, which effectively shuts down its laser ionization. As a consequence, the Rabi flopping dominates, and other complications
are minimized. Below we examine whether we can observe the Autler-Townes doublet for such a "three-level system" where the two
fast-decaying "levels" are strongly coupled by a short pulse.

For the parameters in the model, the $2s^2$ resonance width ($\Gamma_b$) of 0.125~eV, or the lifetime of 5.3~fs, is taken from the
earlier calculation~\cite{chung}. The dipole transition and the $q$-parameter are assumed to be the same as those for $2p^2$, i.e.,
$D_{ba}=2.17$~a.u. and $q_b=1000$, because the first-order transition in $D_{ba}$ is again $\langle 2s|D|2p \rangle$, and $D_{2a}$ has
no first-order term. The PPT rate for $2s^2$ at the laser peak is in the order of 10$^{-5}$~eV which is negligible for our purpose.
The Rabi frequency between $2s2p$ and $2s^2$ is 0.27~eV, corresponding to the period of 15~fs.

\begin{figure}
\centering
\includegraphics[width=0.45\textwidth]{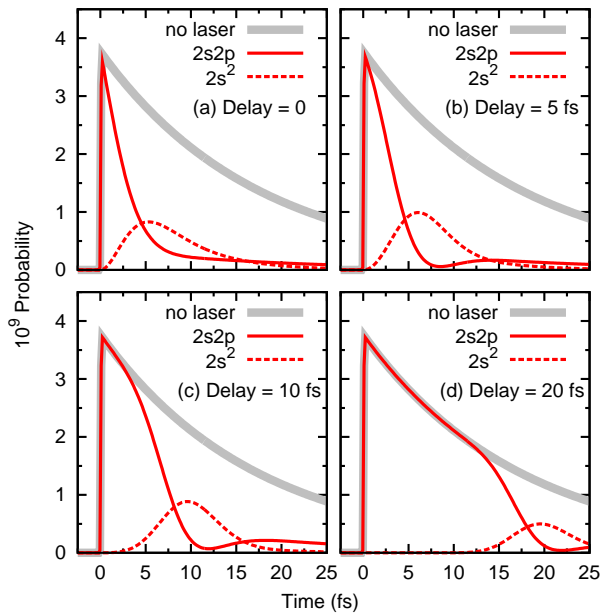}
\caption{(Color online) Probabilities of the $2s2p$ and $2s^2$ bound states with time for $\lambda_L=540$~nm and various delays.
The gray solid curve indicates the laser-free decay of $2s2p$.}
\label{fig-case2-bound}
\end{figure}

The evolution of the bound states is shown in Fig.~\ref{fig-case2-bound}. An obvious difference from the $\lambda_L=780$~nm case is
that the laser coupling now is strong enough to completely deplete $2s2p$ so that its population almost touches zero before bouncing
back, as shown in Figs.~\ref{fig-case2-bound}(b-d). However, when it revives, the amount brought back by the
oscillation is only less than 10\% of what has been removed. For the $2s^2$ state for $t_0=5$, 10, and 20~fs, the overall population
decreases with the delay because the $2s2p$ decays with time and the laser pumps less and less electrons from $2s2p$ to $2s^2$.

\begin{figure}
\centering
\includegraphics[width=0.45\textwidth]{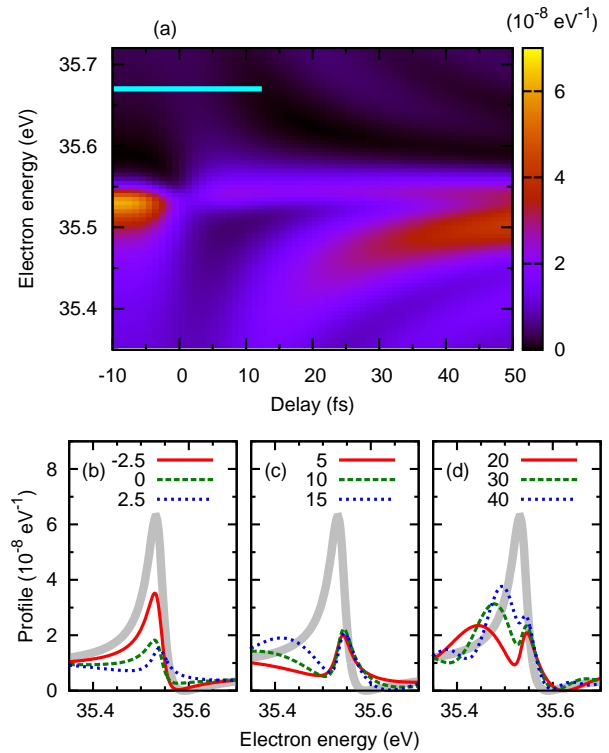}
\caption{(Color online) As Fig.~\ref{fig-case1-a} but for $\lambda_L=540$~nm. The gray solid curve represents the ``no laser'' case.}
\label{fig-case2-a}
\end{figure}

Now we turn to the electron spectra of the $2s2p$ resonance in Fig.~\ref{fig-case2-a}. When the delay passes the origin and becomes
positive, the resonance profile flips horizontally, looking like the mirror image of the ``no laser'' spectrum other than an overall
reduction in height. This flipped peak is seen in the $t_0=5$~fs curve in Fig.~\ref{fig-case2-a}(c). To understand this pattern,
in Fig.~\ref{fig-case2-bound}(b), the sharp drop of $2s2p$ from 0 to 7~fs suggests that most electrons move from $2s2p$ to
$2s^2$ at the very beginning of the decay of $2s2p$. The laser duration allows the Rabi oscillation to run a little more than half
cycle, which bounces only a small fraction of the population back to $2s2p$. The relatively few electrons returning to $2s2p$ have
changed the phase by $\pi$ through this Rabi flopping, thus reversing the sign of $q$ in the resonance profile (See
Ref.~\cite{fano}). When $t_0$ increases, in Fig.~\ref{fig-case2-a}(a), an additional ridge moves from the lower energy to the
original peak position, stretching from 35.4~eV at $t_0=15$~fs to 35.5~eV at $t_0=50$~fs. This suggests that the laser divides the
autoionization into two periods of time; the part prior to the laser is responsible for the regular Fano profile as shown by the ``no
laser'' curve, and the part after the laser is responsible for the ``inverse'' shape that we have just discussed. Different delay
times determine the fractions of these two parts, and the ridge forms at different energies as the interference pattern.
Figure~\ref{fig-case2-a}(d) clearly shows the shifting of the ridge while the inverse peak stays at 35.55~eV.

\begin{figure}
\centering
\includegraphics[width=0.45\textwidth]{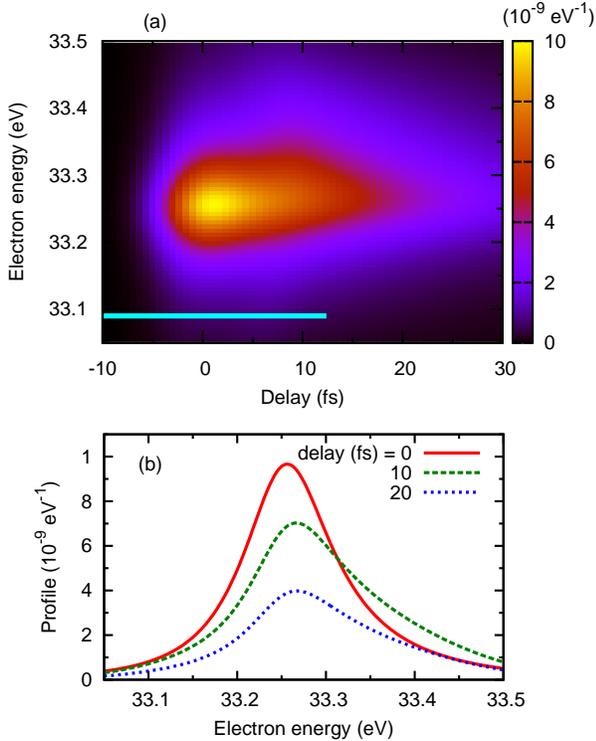}
\caption{(Color online) As Fig.~\ref{fig-case2-a} but for the $2s^2$ resonance.}
\label{fig-case2-b}
\end{figure}

The $2s^2$ profile in Fig.~\ref{fig-case2-b}(a) has a very gentle and monotonic attenuation along $t_0$, while the $2p^2$ profile
in Fig.~\ref{fig-case1-b}(a) features a ``gap'' in $t_0$ where the profile drops to zero altogether before it regains some strength
as $t_0$ increases. This gap, as explained in Sec.~\ref{780nm_cal}, originates from the IR ionization; however in the
$\lambda_L=540$~nm case, this mechanism is missing since the ionization rate for $2s^2$ is low. Rabi oscillation and autoionization
are the only influences on the evolution of $2s^2$. Since the Rabi oscillation runs only a little more than half cycle, it can be
viewed as a ``one way route'' for electrons from $2s2p$ to $2s^2$. Once arriving at $2s^2$, the electrons autoionize quickly
to form the $2s^2$ profile. As $t_0$ increases, more electrons will autoionize from $2s2p$ and less will be brought by the laser to
$2s^2$, and the $2s^2$ resonance becomes weaker.

The choice of $2s^2$ as the second coupled state is optimal for practical measurement issues, such as the signal strength and energy
resolution. One must remember that the $2s^2$ resonance is on top of the $1s\epsilon p$ background if the total electron energies are
measured. The maximum signal intensity of the $2s^2$ profile is nearly 10$^{-8}$~eV$^{-1}$, and the nearby $1s\epsilon p$ background
is $7.5 \times 10^{-9}$~eV$^{-1}$, i.e., when the two are layered together, the resonance signal is high enough to stand out from the
background. Furthermore, its resonance width $\Gamma_b=0.125$~eV is larger than that of $2s2p$, making its visibility better even
with the same energy resolution used in Ref.~\cite{gilbertson}. Considering both the height and the width in spectrum, detecting the
$2s^2$ resonance vs delay should be feasible.

\begin{figure}
\centering
\includegraphics[width=0.45\textwidth]{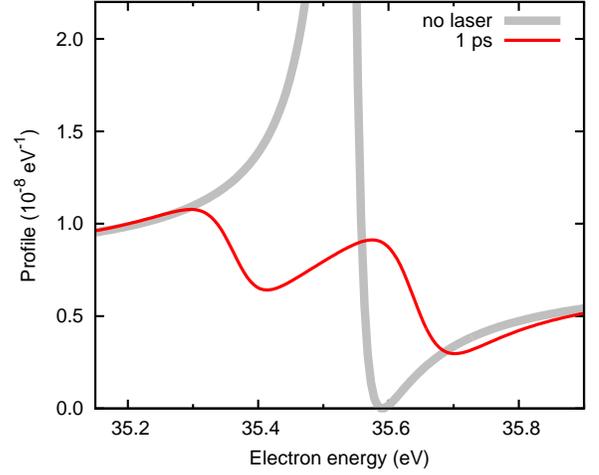}
\caption{(Color online) Photoelectron profile of the $2s2p$ resonance with a 540~nm, 1~ps laser pulse. The original Fano line-shape
is plotted in the gray solid curve.}
\label{fig-eit}
\end{figure}

The complicatedly structured profiles seen in Figs.~\ref{fig-case2-a}(b-d) are in contrast to the Autler-Townes doublet in long-pulse
cases. The latter features the splitting proportional to the field strength, and each peak in the doublet does not differ from
the original single peak in shape. In order to clarify the difference between long and short pulses in a systematic way, in
Fig.~\ref{fig-eit}, we show what the $2s2p$ profile should look like if the laser is 1~ps long, with all other parameters unchanged.
A 1 ps laser pulse is much longer than our atomic time scale and equivalent to a stable AC field, where time delay cannot be defined.
The separation of the splitting is about the Rabi frequency 0.27~eV. Nevertheless, in a standard EIT setup, the $|b\rangle$ state is
a bound state, which is not the
case in the present system. As a result, in our example, the split peaks are fainter because of the finite lifetime of $2s^2$.

If the laser duration shrinks from 1~ps to 50~fs, the magnitude of the splitting will be a function of the time delay.
However, since the laser is still significantly longer than the 17~fs decay lifetime in concern, the
delay-dependence is more relevant to the overlap between the pulses rather than the autoionization. This scenario has been studied
with transient absorption spectroscopy in helium where the XUV is 30~fs and the IR is 42~fs~\cite{loh}. The study showed that
the splitting was tuned to maximum by overlapping the pulses, and it disappeared when the pulses were totally separate.
The system can be viewed as a dressed atom where the dressing condition can be slowly turned on or off by changing the time delay.

As the comparison indicates, the mechanisms for the short and long pulses are indeed very different. For a long
dressing pulse, the field enters as the coupling (off-diagonal) terms in the Hamiltonian of the system. When the Hamiltonian is
diagonalized, the energy levels are shifted, which are then represented by the doublet. However, for a short dressing pulse, the
coupling strength changes quickly; the Rabi frequency is not well-defined, and the mechanism for the long pulse breaks down.
In other words, the dressing field influences only a short temporal segment out of the whole autoionization, while most of the time,
the resonance profile evolves without laser and aims at the single-peak Fano profile rather than the doublet.

\section{Summary and Conclusions}\label{conclusions}

A model has been developed to study the autoionization dynamics in a laser-dressed helium. An attosecond XUV excites the $2s2p$
resonance in a time-delayed 780~nm IR pulse. The IR can couple the $2s2p$ and $2p^2$ states and ionize the two states. The
photoelectron energy spectra for different time delays are calculated and compared with the experiment~\cite{gilbertson}. While the
experimental energy resolution was not good enough to observe the resonance shape in detail, good agreement for the resonance peak
intensity vs time delay between our model and the experiment has been achieved. The decay lifetime of $2s2p$ can be retrieved by this
result. Because of the strong IR ionization, the coupling with $2p^2$ is to open an efficient pathway where the $2s2p$ resonance 
can be depleted, i.e., the IR field modifies the profile with an overall depression without changing the spectral shape, which is
totally different from the typical three-level systems.

To reduce the effect of ionization by the IR laser, we change the laser wavelength to 540~nm and consider the coupling of the $2s2p$
and $2s^2$ states. The $2s^2$ state has a larger binding energy where its ionization is negligible. In this case, a complicated
pattern in the $2s2p$ resonance shape vs time delay has been found. The result is interpreted by the Rabi oscillation between the two
autoionizing states whose cycles are confined by the 9~fs laser. In order to make connection of this result to the traditional
dressed-atoms, we change the laser duration to 1~ps and recover the Autler-Townes splitting in the spectrum, which also clarifies the
difference between the short- and long-dressing fields. In order words, in the presence of an intense IR, autoionization dynamics can
be changed significantly. The possibilities of such manipulations are tremendous, and what one can gain from such experiments remains
to be further explored.

\begin{acknowledgments}
This work was supported in part by Chemical Sciences, Geosciences and Biosciences Division, Office of Basic Energy
Sciences, Office of Science, U.S. Department of Energy.
\end{acknowledgments}

\end{document}